\documentclass[10pt]{article}
\usepackage{amssymb}
\newtheorem{thm}{Theorem}[section]
\newtheorem{prop}[thm]{Proposition}
\newtheorem{lem}[thm]{Lemma}

\newcommand{\BZ}{{\mathbb Z}}

\newcommand{\BP}{{\mathbb P}}

\newcommand{\BC}{{\mathbb C}}

\newcommand{\qed}{\hfill$\square$}
\newcommand{\E}{e}
\newcommand{\ep}{{\varepsilon}}

\newcommand{\dfrac}[2]{{\displaystyle\frac{#1}{#2}}}
\newcommand{\ol}{\overline}
\newcommand{\ul}{\underline}
\def\7b{{\overline 7}}
\def\6u{{\underline 6}}
\newcommand{\oy}{\overline {y}}
\newcommand{\of}{\overline {f}}
\newcommand{\ux}{\underline {x}}
\newcommand{\uf}{\underline {f}}
\def\ll{{\ell}}
\begin{document}
\begin{center}
{\Large
${}_{10}E_9$ solution to the elliptic Painlev\'e equation
}
\vskip5mm
{\normalsize\em Dedicated to Professor Kyoichi Takano on his sixtieth birthday}
\vskip10mm
{\large 
Kenji Kajiwara\footnote{
Graduate School of Mathematics, Kyushu University,
Hakozaki, Fukuoka 812-8512, Japan.},
Tetsu Masuda\footnote{
Department of Mathematics, Kobe University,
Rokko, Kobe 657-8501, Japan.},
Masatoshi Noumi$^2$,\\[2mm]
Yasuhiro Ohta$^2$ 
and Yasuhiko Yamada$^2$}
\end{center}

\vskip10mm

\begin{abstract}
A $\tau$ function formalism for Sakai's elliptic Painlev\'e equation is
presented. This establishes the equivalence between the two formulations by 
Sakai and by Ohta-Ramani-Grammaticos.
We also give a simple geometric description of the elliptic
Painlev\'e equation as a non-autonomous
deformation of the addition formula on elliptic curves.  
By using these formulations, we construct a particular solution 
of the elliptic Painlev\'e equation
expressed in terms of the elliptic hypergeometric function
${}_{10}E_{9}$.
\end{abstract}

\section{Introduction}
The elliptic Painlev\'e equation proposed by Sakai\cite{Sakai} is located 
on the top of Painlev\'e and discrete Painlev\'e equations. 
It is a second order nonlinear difference
equation with affine Weyl group symmetry of type $E_8^{(1)}$, and
regarded as the discrete dynamical system on a rational surface obtained 
by blowing-up at nine points on $\BP^2$. 
All the Painlev\'e and discrete Painlev\'e equations
are derived by degeneration from the elliptic Painlev\'e equation.

On the other hand, it is well-known that Painlev\'e and discrete
Painlev\'e equations admit special cases which are reducible to the
Riccati equations. From them we obtain particular solutions described
by special functions of hypergeometric type\cite{GtoP,RGTT,MSY}. 
We call such solutions the {\em Riccati solutions} or the 
{\em hypergeometric solutions}. 
A natural question is what kind of function of hypergeometric 
type arises as such a particular solution to the elliptic Painlev\'e equation. 
The purpose of this letter is to present an answer to this question: 
it is the elliptic hypergeometric function ${}_{10}E_9$ introduced by 
Frenkel-Turaev\cite{FT}, and studied further 
by Spiridonov-Zhedanov\cite{SZ}.

\par\medskip
In Section 2, we introduce a framework of $\tau$ functions for the elliptic 
Painlev\'e equation, and describe explicitly the birational action of the affine 
Weyl group $W(E^{(1)}_{8})$ on the level of $\tau$ functions. 
We also present bilinear equations to be satisfied 
by the $\tau$ functions. 
In Section 3, we give an explicit form of the elliptic Painlev\'e equation, 
and construct a particular solution described by ${}_{10}E_9$. 
We also present a simple geometric description for
the elliptic Painlev\'e equation in Section 4, 
and give another method for constructing the hypergeometric solution. 

\section{$\tau$ function}

Let us recall briefly Sakai's formulation of the elliptic Painlev\'e 
equation\cite{Sakai}.  This approach, based on the geometry of rational 
surfaces, was initiated by Okamoto\cite{O} and pursued further by 
Takano (see for instance \cite{ST}). 
Consider the configuration space ${\cal M}_{m,n}$ ($0<m<n$) of $n$ points
$P_i=(x_{1i}:x_{2i}:\ldots:x_{mi})$ ($1\leq i \leq n$) on $\BP^{m-1}$:
\begin{equation}
{\cal M}_{m,n}={\rm GL}(m) \backslash
\left\{
\left[\begin{array}{ccccc}
x_{11}&x_{12}&\cdots&x_{1n}\\
x_{21}&x_{22}&\cdots&x_{2n}\\
\vdots&\vdots&&\vdots\\
x_{m1}&x_{m2}&\cdots&x_{mn}
\end{array}\right]
\right\}
/ (\BC^{\times})^{n}.
\end{equation}
Let $W_{m,n}$ be the Weyl group generated by simple reflections
$s_i$ ($i=0,1,\ldots,n-1$) corresponding to the following Dynkin-Coxeter
diagram:

$$
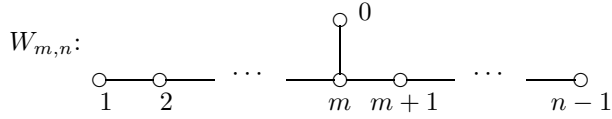
\setlength{\unitlength}{0.8mm}
\begin{picture}(100,30)(-20,-10)
\put(0,-5){1}
\put(10,-5){2}
\put(38,-5){$m$}
\put(45,-5){$m+1$}
\put(75,-5){$n-1$}
\put(43,10){0}
\put(-15,5){$W_{m,n}$:}
\put(0,0){\circle{2}}
\put(10,0){\circle{2}}
\put(22,0){$\cdots$}
\put(40,0){\circle{2}}
\put(50,0){\circle{2}}
\put(62,0){$\cdots$}
\put(80,0){\circle{2}}
\put(40,10){\circle{2}}
\put(1,0){\line(1,0){8}}
\put(11,0){\line(1,0){8}}
\put(31,0){\line(1,0){8}}
\put(41,0){\line(1,0){8}}
\put(51,0){\line(1,0){8}}
\put(71,0){\line(1,0){8}}
\put(40,1){\line(0,1){8}}
\end{picture}
$$
There is a birational action of the Weyl group $W_{m,n}$ on the
configuration space ${\cal M}_{m,n}$ \cite{OD}.  
In the case of $(m,n)=(3,10)$, the Weyl group $W_{3,10}$ contains 
a translation subgroup $\BZ^8 \subset W(E^{(1)}_8)=W_{3,9} \subset W_{3,10}$.
The birational action of the translation is nothing but
the elliptic Painlev\'e equation\cite{Sakai}; the points $P_1,
\ldots, P_9$ play the role of parameters, while the last point $P_{10}$
is the dependent variable \cite{MY}.

Let $P=\BZ \E_0 \oplus \BZ \E_1 \oplus \cdots 
\oplus \BZ \E_n$ be the lattice with the metric defined by 
\begin{equation}
\Big\langle \sum_{i=0}^n x_i \E_i, \sum_{i=0}^n y_i \E_i \Big\rangle=
-(m-2)x_0y_0+\sum_{i=1}^n x_i y_i.
\end{equation}
We consider the root lattice
$Q=\BZ \alpha_0 \oplus \cdots \oplus \BZ \alpha_{n-1} \subset P$
generated by 
\begin{equation}
\alpha_i=\E_i-\E_{i+1}, \quad (1 \leq i \leq n-1), \quad
\alpha_0=\E_0-\E_1-\E_2-\cdots -\E_m.
\end{equation}
The standard formula of the reflection with respect to $\alpha_i$ 
($i=0,\ldots,n-1$),
\begin{equation}\label{WeylPic}
s_i(L)=L-\langle \alpha_i,L \rangle \alpha_i,\quad (i=0,\ldots,n-1)
\end{equation}
gives the action of the Weyl group $W_{m,n}$ on the lattice $P$.
In the Painlev\'e context, vectors $v \in \sum_{i=0}^n \BC \E_i$ play
the role of parameters (or independent variable), whose coordinates are 
denoted by $\ep_i=\langle \E_i, v \rangle $ ($i=0,\ldots,n$).
Then $W_{m,n}$ acts on them as follows:
\begin{equation}
\begin{array}l
s_0(\ep_0)=(m-1)\ep_0-\ep_1-\cdots-\ep_m, \\[2mm]
s_0(\ep_i)=(m-2)\ep_0-\ep_1-\cdots-\widehat \ep_i-\cdots-\ep_m, 
\quad (i=1,\ldots,m), \\[2mm]
s_0(\ep_j)=\ep_j, \quad (j=m+1,\ldots,n), \\[2mm]
s_i(\ep_i)=\ep_{i+1}, \quad
s_i(\ep_{i+1})=\ep_{i}, \quad (i=1,\ldots,n-1),\\[2mm]
s_i(\ep_j)=\ep_j \quad (\mbox{otherwise}).
\end{array}
\end{equation}
Let $K$ be the field of entire meromorphic functions in $(\ep_0,\ep_1,\ldots,\ep_n)$, and 
consider the field $K(f)$ of 
rational functions in the indeterminates $f=(f_1,\ldots,f_m)$ with 
coefficients in $K$. 
Let ${\cal R}=K(f)[\tau_1^{\pm},\ldots,\tau_n^{\pm}]$ 
be the ring of Laurent
polynomials in $(\tau_1,\ldots,\tau_n)$ 
over $K(f)$. 
Then one can construct a birational representation of
$W_{m,n}$ on ${\cal R}$ as follows:
\begin{thm}
We define the birational actions of $s_i$ $($ $i=0,\ldots,n-1$ $)$ 
on ${\cal R}$ as follows:
\begin{equation}\label{birational-tau:mn}
\begin{array}l
s_0(\tau_i)=f_i \tau_{i}, \quad (i=1,\ldots,m), \quad
s_0(\tau_j)=\tau_j, \quad (j=m+1,\ldots,n), \\[2mm]
s_i(\tau_i)=\tau_{i+1}, \quad s_i(\tau_{i+1})=\tau_i, \quad(i=1,\ldots,n-1),\\[2mm]
s_i(\tau_j)=\tau_j,\quad (\mbox{otherwise}),
\end{array}
\end{equation}
\begin{equation}\label{birational-f:mn}
\begin{array}l
\smallskip
s_0(f_i)=\dfrac{1}{f_i}, \quad (i=1,\ldots,m),\\[2mm]
\smallskip
s_i(f_i)=f_{i+1}, \quad s_i(f_{i+1})=f_i, \quad(i=1,\ldots,m-1),\\[2mm]
\smallskip
s_m(f_i)=\dfrac{\tau_m}{\tau_{m+1}}\left(c_{i,m}f_i+c_{m,i}f_m\right), \quad (i=1,\ldots,m-1),\\[2mm]
s_m(f_m)=\dfrac{\tau_m}{\tau_{m+1}}f_m,\\[2mm]
s_i(f_j)=f_j,\quad (\mbox{otherwise}),
\end{array}
\end{equation}
where $c_{i,j}=
\dfrac{[\ep_i-\ep_{m+1}][\alpha_0+\ep_j-\ep_{m+1}]}
{[\ep_i-\ep_j][\alpha_0]}$ and $[x]=\vartheta_{11}(x)$ is the odd theta
function. Then $\langle s_0,\ldots,s_{n-1}\rangle$ forms the Weyl group $W_{m,n}$.
\end{thm}

\noindent
{\it Proof}. 
By direct computation, the Coxeter relation $(s_is_j)^2=1$ or
$(s_is_j)^3=1$ follows from the Riemann relation
\begin{equation}
[a+b][a-b][c+z][c-z]+(abc \ {\rm cyclic})=0,
\end{equation}
and $[-x]=-[x]$.\qed
\par\medskip
Let $M$ be the Weyl group orbit $M=W_{m,n}.\E_1 \subset P$.
For any $\Lambda \in M$, choose $w \in W_{m,n}$ such that
$w.\E_1=\Lambda$ and define the $\tau$ function $\tau(\Lambda)$ as
\begin{equation}\label{tau-general}
\tau(\Lambda)=w(\tau_1).
\end{equation}
This definition is independent of the choice of $w$ and we have
\begin{equation}
\begin{array}l
w(\tau(\Lambda))=\tau(w.\Lambda), 
\quad (w \in W_{m,n}, \ \Lambda \in M),\\[2mm]
\tau(\E_i)=\tau_i, \quad (i=1,\ldots,n).
\end{array}
\end{equation}
Then, one can derive the following bilinear difference equations of
Hirota-Miwa type for the $\tau$ functions:
\begin{prop}\label{bilinear-general}
For any mutually distinct indices
$1 \leq i,j,k,l_1,\ldots,l_{m-2} \leq n$, we have
\begin{equation}\label{bl:mn}
\begin{array}{l}
\phantom{+}
[\ep_j-\ep_k][L-\ep_j-\ep_k]\tau(\E_i)\tau(L-\E_i) \\[2mm]
+\,[\ep_k-\ep_i][L-\ep_k-\ep_i]\tau(\E_j)\tau(L-\E_j) \\[2mm]
+\,[\ep_i-\ep_j][L-\ep_i-\ep_j]\tau(\E_k)\tau(L-\E_k)=0,
\end{array}
\end{equation}
where $L=\E_0-\E_{l_1}-\cdots-\E_{l_{m-2}}$. 
\end{prop}

\noindent
{\it Proof.} Using the relation $f_i=s_0(\tau_i)/\tau_i$, 
the formula for $s_m(f_i)$ can be written in terms of $\tau$ as
\begin{equation}\label{tsst}
\tau_{m+1}s_m s_0(\tau_i)=c_{i,m}\tau_m s_0(\tau_i)+c_{m,i}\tau_i s_0(\tau_m).
\end{equation}
This equation is a special case of eq.(\ref{bl:mn}) where
$(i,j,k)=(i,m,m+1)$ and $L=\alpha_0+\E_i+\E_m$.
Other cases can be obtained by the ${\mathfrak S}_n$-symmetry.\qed
\par\medskip
Moreover, by reversing the above argument, we have the following:
\begin{thm}
For any family of functions $\tau(\Lambda)$ $(\Lambda \in M)$, we define
variables $f_i$ by $f_i=\tau(s_0.\E_i)/\tau(\E_i)$ for
$i=1,\ldots,m$.  Then the actions of the Weyl group $W_{m,n}$ on $M$ are
consistent with the transformations (\ref{birational-tau:mn}) and
 (\ref{birational-f:mn}) of $f_i$ and $\tau_j=\tau(\E_j)$
$(i=1,\ldots,m$, $j=1,\ldots, n)$ if and only if the bilinear relations
(\ref{bl:mn}) are satisfied.
\end{thm}

In what follows, we consider the case $(m,n)=(3,9)$ of nine points 
$P_1,\ldots,P_9$ in $\BP^2$, together with an additional point 
$P_{10}$ playing the role of the dependent variable of 
the elliptic Painlev\'e equation.  
The formulation of $\tau$ functions described above 
is relevant to the following parametrization of the 
elliptic curve $C_0$ passing through the nine points:
\begin{equation}\label{eq:Pu}
P(u)=\Big(\dfrac{[\ep_0-\ep_2-\ep_3-u]}{[\ep_1-u]}:
\dfrac{[\ep_0-\ep_3-\ep_1-u]}{[\ep_2-u]}:
\dfrac{[\ep_0-\ep_1-\ep_2-u]}{[\ep_3-u]}\Big).
\end{equation}
The nine points $P_i$ on $C_0$ are given by 
$P_i=P(\ep_i)$ ($i=1,\ldots,9$).
The variables $(f_1:f_2:f_3)$ then reprensent the 
homogeneous coordinates of the last point $P_{10}$.

The lattice $P$ is interpreted as the Picard lattice of the
rational surface $X$ obtained by blowing-up of $\BP^2$ at the 
nine points $P_i$ ($i=1,\ldots,9$). 
The metric $\langle \alpha,
\beta \rangle$ is minus the intersection pairing. 
The null vector $\delta$ defined by $\delta=3 \E_0-\sum_{i=1}^9 \E_i$
represents the anti-canonical divisor of $X$.
An element $L \in P$ defines a linear system $|L|$ on $\BP^2$.
For $L=d \E_0-\sum_{i=1}^9 m_i \E_i \in P$, the linear system $|L|$, 
classically denoted by $C^d(P_1^{m_1}\cdots P_9^{m_9})$ \cite{SR},
represents a complete family of curves in $\BP^2$ determined as the zero 
locus of the homogeneous polynomial with assigned degree
$d$ and zero multiplicity $m_i$ at $P_i$ ($i=1,\ldots,9$). 
The virtual genus $g(L)$ of a curve $C \in |L|$ and the 
virtual dimension ${\rm dim}\,\vert L \vert$ of the family $|L|$
is given by
\begin{equation}\label{gandL}
2-2g(L)=\langle L,L-\delta \rangle, \quad
2~{\rm dim}\,\vert L \vert =-\langle L,L+\delta \rangle,
\end{equation}
respectively. 
In the situation here, the above formulas give the real genus 
and dimension, respectively\cite{Nagata}.

In our formulation, the Weyl group orbit $M=W(E^{(1)}_8).\E_1 \subset P$ 
plays the essential role.
The formulas eq.(\ref{gandL}) imply that $g(\Lambda)=0$ and 
${\rm dim}\,|\Lambda|=0$  for any $\Lambda \in M$. 
It then turns out that the $\tau$ function $\tau(\Lambda)$ 
is a homogeneous polynomial of degree $d$ in the variables $(f_1,f_2,f_3)$.
Furthermore, the equation $\tau(\Lambda)=0$ specifies the unique 
rational curve corresponding to the linear system associated with 
$\Lambda \in M$. 
Due to our normalization of the $\tau$ functions $\tau(\Lambda)$ 
in eqs.(\ref{birational-tau:mn}) and (\ref{tau-general}), we have
\begin{equation}\label{tau-normalization}
\tau(\Lambda) \Big|_{C_0}=
[\Lambda-u]\prod_{i=1}^9 [\ep_i-u]^{m_i}\,
\tau_0^{d}\,\prod_{i=1}^9 \tau_i^{-m_i}
\qquad
(\tau_0=\tau_1\tau_2\tau_3), 
\end{equation}
for $\Lambda=d \E_0-\sum_{i=1}^9 m_i \E_i \in M$,
under the parametrization eq.(\ref{eq:Pu}) of $C_0$. 
We note that for a generic curve $C_m$ of degree $m$, 
the $3m$ intersecting points
$P(u_i) \in C_m \cap C_0$ ($i=1,2,\ldots,3m$) satisfy the relation 
$[m \ep_0-\sum_{i=1}^{3m} u_i]=0$. 

As a typical case of Proposition \ref{bilinear-general}, 
our $\tau$ functions satisfy the following bilinear relation:
\begin{equation}\label{GRO=S}
\begin{array}l
\phantom{+}
[\ep_2-\ep_3][\ep_0-\ep_2-\ep_3-\ep_4] 
\tau(\E_1)\tau(\E_0-\E_1-\E_4) \\[2mm]
+\,[\ep_3-\ep_1][\ep_0-\ep_3-\ep_1-\ep_4] 
\tau(\E_2)\tau(\E_0-\E_2-\E_4) \\[2mm]
+\,[\ep_1-\ep_2][\ep_0-\ep_1-\ep_2-\ep_4] 
\tau(\E_3)\tau(\E_0-\E_3-\E_4)=0.
\end{array}
\end{equation}
This formula can also be proved geometrically 
as follows.
For $L=\E_0-\E_4$, we have ${\rm dim}\,|L|=1$. 
Hence there must be a linear relation among the three terms. The coefficients 
can be determined by the normalization condition eq.(\ref{tau-normalization}).

\par\medskip
In the paper \cite{RGO}, a bilinear formulation of the elliptic Painlev\'e
equation with $W(E^{(1)}_8)$ symmetry was proposed.
Written in terms of $E_8$ lattice, our bilinear relations coincide
with those in \cite{RGO}.
This fact establishes equivalence
between the two formulations of the elliptic Painlev\'e equation.

\section{Hypergeometric solutions}
In order to determine the explicit form of the elliptic Painlev\'e equation, 
we consider the actions of translations in $W(E_8^{(1)})$. 
\begin{lem}$\cite{Kac}$
For a translation $T_{\alpha} \in W(E^{(1)}_8)$ 
along a vector in the root lattice $\alpha \in Q$, one has
\footnote{A relation between the coefficient of $e_0$ and the algebraic
entropy is pointed out in ref.\cite{Take}.}
\begin{equation}
T_{\alpha}(L)=L+k \alpha-\left(
\dfrac{k}{2}\langle \alpha, \alpha \rangle+\langle L,\alpha \rangle
\right )\delta, \quad
k=\langle \delta,L \rangle.
\end{equation}
\end{lem}

The 240 roots of $E_8$ are represented as 
$\alpha_{ij}=\E_i-\E_j$ ($72$ vectors) or
$\pm \alpha_{ijk}=\pm(\E_0-\E_i-\E_j-\E_k)$ ($2 \times 84$ vectors).
The corresponding translations are given by
\begin{equation}
\begin{array}{ll}
T_{\alpha_{ij}}=s_{ia_1a_2}s_{ia_3a_4}
s_{a_5a_6a_7}s_{ia_3a_4}s_{ia_1a_2}s_{ij},&
\{i,j,a_1,\ldots,a_7\}=\{1,2,\ldots,9\},\\[3mm]
T_{\alpha_{ijk}}=s_{a_1a_2a_3}s_{a_4a_5a_6}s_{a_7a_8a_9}s_{ijk},
&\{i,j,k,a_1,\ldots,a_6\}=\{1,2,\ldots,9\},
\end{array}
\end{equation}
where $s_{ij}$ and $s_{ijk}$ are the reflections with respect to the
roots $\alpha_{ij}$ and $\alpha_{ijk}$ respectively. 
In what follows, we consider only the translations of type $T_{\alpha_{ij}}$.
As an example, let us consider the translation 
\begin{equation}
T_{\alpha_{6}}=s_{126}s_{346}s_{589}s_{346}s_{126}s_{67}.
\end{equation}
To write down the explicit formula of the elliptic Painlev\'e equation 
with respect to the translation $T_{\alpha_6}$, it is convenient to factorize
the $T_{\alpha_6}$ as
\begin{equation}
T_{\alpha_6}={\mu}^2, \quad \mu=s_{126}s_{346}s_{568}s_{15}s_{28}s_{79}s_{67}.
\end{equation}
The actions of $\mu$ on the parameters $\ep_i$ are given by
\begin{equation}\label{muep}
\begin{array}{ll}
\mu(\ep_0)=\delta+\ep_0-2 \ep_6+\ep_7+\ep_9, &
\mu(\ep_6)=\ep_9, \\[2mm]
\mu(\ep_7)=\delta-\ep_6+\ep_7+\ep_9, &
\mu(\ep_9)=\ep_7, \\[2mm]
\mu(\ep_i)=\ep_0-\ep_6-\ep_j,
\end{array}
\end{equation}
where, in the last equation, 
$(i,j)=(1,8),(2,5),(3,4),(4,3),(5,2),(8,1)$. 
Here, by abuse of notation, we denote $\delta=3 \ep_0-\sum_{i=1}^9 \ep_i$.
This $\delta$ appears only in the theta function $[x]$ and plays the
role of the unit for the elliptic difference operation.
We introduce intermediate variables $g_i$ ($i=1,2,3$) by
\begin{equation}\label{mug}
g_i=\mu(f_i)=\dfrac{\tau(\mu s_0.\E_i)}{\tau(\mu.\E_i)}
=\dfrac{\tau(2\E_0-\E_1-\E_2-\E_3-\E_6-\E_j)}{\tau(\E_0-\E_6-\E_j)},
\end{equation}
where $(i,j)=(1,8),(2,5),(3,4)$. 
By a similar method in the proof
of eq.(\ref{GRO=S}),
one can show the following formulas for $i,j \geq 4$ $(i\neq j)$:
\begin{equation}\label{mutau}
\begin{array}l
\tau(\E_0-\E_i-\E_j)\tau_i\tau_j
=\dfrac{[1i][1j][1ij]}{[12][13][123]}\tau(\E_0-\E_2-\E_3)\tau_2\tau_3+
{(123 \ \rm cyclic)}, \\[4mm]
\tau(\E_0-\E_1-\E_2-\E_3-\E_i-\E_j)\tau_i\tau_j\\[4mm]
=-\dfrac{[23i][23j][1ij]}{[12][13][123]}\tau_1\tau(\E_0-\E_1-\E_3)
\tau(\E_0-\E_1-\E_2)+{(123 \ \rm cyclic)},
\end{array}
\end{equation}
where $[ij]=[\ep_i-\ep_j]$ and $[ijk]=[\ep_0-\ep_i-\ep_j-\ep_k]$.
From these formulas we obtain
\begin{prop}\label{prop:gTf}
The action of $T_{\alpha_6}$ is given explicitly in the form
\begin{equation}\label{gTf}
g_i=\frac{Q_i(f)}{P_i(f)}, \qquad
\ol{f_i}=T_{\alpha_6}(f_i)=\frac{S_i(g)}{R_i(g)}\qquad(i=1,2,3).
\end{equation}
The polynomials $P_i,Q_i,R_i$ and $S_i$ are given by
\begin{equation}
\begin{array}{l}
P_i=
 \dfrac{[1j][16][1j6]}{[12][13][123]}f_1
-\dfrac{[2j][26][2j6]}{[12][23][123]}f_2
+\dfrac{[3j][36][3j6]}{[13][23][123]}f_3, \\[4mm]
Q_i=
-\dfrac{[23j][236][1j6]}{[12][13][123]}f_2f_3
+\dfrac{[13j][136][2j6]}{[12][23][123]}f_1f_3
-\dfrac{[12j][126][3j6]}{[13][23][123]}f_1f_2, \\[4mm]
R_i=
-\dfrac{[i8][689][i\7b 8]}{[48][58][123]}g_1
+\dfrac{[i5][569][i5\7b ]}{[45][58][123]}g_2
-\dfrac{[i4][469][i4\7b ]}{[45][48][123]}g_3, \\[4mm]
S_i=
-\dfrac{[ab8][45\7b ][i\7b 8]}{[48][58][123]}g_2g_3
+\dfrac{[ab5][4\7b 8][i5\7b ]}{[45][58][123]}g_1g_3
-\dfrac{[ab4][5\7b 8][i4\7b ]}{[45][48][123]}g_1g_2,
\end{array}
\end{equation}
where $j=8,5,4$ and $(a,b)=(2,3),(1,3),(1,2)$ for $i=1,2,3$,
respectively, and
$\ep_{\7b}=\ep_7+\delta$.
\end{prop}
\noindent
{\it Proof.} The first equation of eq.(\ref{gTf}) is a consequence
of eqs.\,(\ref{mug}) and (\ref{mutau}). 
The second equation is obtained by applying
$\mu$ on the first one, namely, by replacing $f_i$ with $g_i$, $g_i$ with
$\ol{f_i}$, and $\ep_i$ with $\mu(\ep_i)$ of eq.(\ref{muep}).\qed
\par\medskip
Notice that the system of difference equations eq.(\ref{gTf}) is 
written in terms of the homogeneous coordinates of $\BP^2$.  
This system, apparently of third order, gives rise to a system of 
second order in the inhomogeneous coordinates. 

\par\medskip
To obtain the hypergeometric solution, let us put
$[34]=0$. Then we may consistently specialize the variables 
so that $\tau_3=0$, $f_3=\infty$ and $g_3=\infty.$\footnote{
Another possibility is the case of $[124]=0$
where the specialization $f_3=g_3=0$ is available,
which is the $s_0$ transform of $[34]=0$, $f_3=g_3=\infty$.
More generally, if $[ijk]=0$ one can consistently specialize as
$\tau(\E_0-\E_i-\E_j)=\tau(\E_0-\E_j-\E_k)=\tau(\E_0-\E_k-\E_i)=0$.
We will discuss the latter
case in the next section with a different method.}
Then we have
\begin{equation}
\begin{array}{l}
\ol{f_i}=\dfrac{-[38][ab5][3\7b 8][i5\7b ]g_1+[35][ab8][35\7b ][i\7b 8]g_2}
{[58][i3][369][i3\7b ]}, \\[4mm]
g_i=\dfrac{[13][13j][136][2j6]f_1-[23][23j][236][1j6]f_2}{[12][3j][36][3j6]}, 
\end{array}
\end{equation}
for $i=1,2$, with $(a,b)$ and $j$ as in Proposition \ref{prop:gTf}. 
Eliminating the variables $g_i$, we have
\begin{equation}\label{olfi}
\ol{f_i}=\frac{a_i f_1+b_i f_2}{c_i},\qquad (i=1,2),
\end{equation}
where
\begin{equation}
\begin{array}l
a_i=[13][136]
(-[ab5][3\7b 8][i5\7b ][356][138][268]
+[ab8][35\7b ][i\7b 8][368][135][256]),\\[2mm]
b_i=[23][236]
([ab5][3\7b 8][i5\7b ][356][238][168]
-[ab8][35\7b ][i\7b 8][368][235][156]),\\[2mm]
c_i=[58][i3][369][i3\7b ][12][36][356][368].
\end{array}
\end{equation}
{}From eq.\,(\ref{olfi}) and its $T_{\alpha_6}^{-1}$ transformation
(obtained by replacing $f_i$, $\ol{f_i}$, $\ep_{\7b}$ and $\ep_6$ with 
$\ul{f_i}$, $f_i$, $\ep_7$ and $\ep_{\6u}=\ep_6+\delta$, respectively),
we further eliminate $f_2$ and obtain a three-term relation
for $\ol{f_1}$, $f_1$ and $\ul{f_1}$.
We will show that this three-term relation can be identified with 
the difference equation for
the elliptic hypergeometric function ${}_{10}E_9$.
To this end, we apply the transformation $s_{289}s_{48}s_{39}$.
Then we have the second order equation for 
$F=s_{289}s_{48}s_{39}(f_1)=\tau_8/\tau_1$, 
under the specialization $[89]=0$.
Explicitly, it is given by
\begin{equation}
\ol{F}=\dfrac{A_1}{A_2} F-\dfrac{B_1}{B_2} \ul{F}, 
\end{equation}
where 
\begin{equation}
\begin{array}l
A_1=[79][169][\6u 7]\\
\phantom{A_1=}\times(-[59][4\7b 9][246][346][149][569]+[49][5\7b 9][256][356][159][469])\\
\phantom{A_1}
+[69][179][6\7b ]\\
\phantom{+[69]}\times([49][5\6u 9][257][357][159][479]-[59][4\6u 9][247][347][149][579]),\\[2mm]
\displaystyle
A_2=[199][45][\6u 7][1\7b 9][79] \prod_{k=2}^5[k69],\\[2mm]
\displaystyle
B_1=[6\7b ][1\6u 9][169][69]\prod_{k=1}^5[k79], \quad
B_2=[\6u 7][1\7b 9][179][79]\prod_{k=1}^5[k69].
\end{array}
\end{equation}
Further, we apply a gauge transformation $F=g\,\Phi$ by using 
a solution $g$ of the difference equation
\begin{equation}
\ol{g}=\frac{[169][\7b 99]}{[1\7b 9][699]}g.
\end{equation}
Then we finally obtain the following: 
\begin{thm}
The linear difference equation for $\Phi$ is determined as 
\begin{eqnarray}\label{eq:Phi}
&&\frac{[79][\7b 99][799]\prod_{k=1}^5[k69]}{[76][\7b 6]}
\left(\ol{\Phi}-\Phi\right)\nonumber\\
&=&\frac{[69][\6u 99][699]\prod_{k=1}^5[k79]}{[76][7\6u]}
\left(\Phi-\ul{\Phi}\right)
-[679]\prod_{k=1}^5[k9]\ \Phi. 
\end{eqnarray}
\end{thm}
We introduce the parameters $u_i$ ($i=0,1,\ldots,7$) by setting 
\begin{equation}
u_0=\ep_0-3 \ep_9-\delta,\qquad
u_i=\ep_i-\ep_9,\quad(i=1,\ldots,7). 
\end{equation}
Note that the parameters $u_i$ satisfy the balancing restriction
\begin{equation}
2 \delta+3 u_0-\sum_{i=1}^7 u_i=0. 
\end{equation}
Then the differece equation eq.(\ref{eq:Phi}) for $\Phi$ can 
be rewritten into the {\em elliptic hypergeometric equation}
\begin{equation}\label{10E9-eq}
\begin{array}l
\dfrac{[u_7][u_0-u_7][\delta+u_0-u_7]\prod_{i=1}^5[\delta+u_0-u_i-u_6]}
{[u_7-u_6][u_7-u_6+\delta]}\Big[\Phi(6^{-}7^{+})-\Phi\Big] \\[4mm]
=\dfrac{[u_6][u_0-u_6][\delta+u_0-u_6]\prod_{i=1}^5[\delta+u_0-u_i-u_7]}
{[u_7-u_6][u_7-u_6-\delta]}\Big[\Phi-\Phi(7^{-}6^{+})\Big] \\[4mm]
+\prod_{i=1}^5[u_i][u_7+u_6-u_0-\delta]\Phi,
\end{array}
\end{equation}
where $\Phi(i^{-}j^{+})=
\Phi \vert_{u_i \mapsto u_i-\delta, u_j \mapsto u_j+\delta}$.
It is known \cite{SZ} that if one of the parameters $u_i$ ($i=1,\ldots,7$) 
is $0,-\delta, -2 \delta, \cdots$, then the terminating very-well-poised
balanced elliptic hypergeometric series 
\begin{equation}
\begin{array}c
\displaystyle
\Phi={}_{10}E_{9}(u_0,u_1,\ldots,u_7)
=\sum_{n=0}^{\infty}\dfrac{[u_0+2n \delta]}{[u_0]}\prod_{r=0}^5
\dfrac{[u_r]_n}{[u_0-u_r+\delta]_n},\\[4mm]
[z]_n=[z][z+\delta]\cdots[z+(n-1)\delta],
\end{array}
\end{equation}
solves the linear equation eq.(\ref{10E9-eq}).
Hence, the elliptic hypergeometric function 
${}_{10}E_9$ gives a particular solution of the elliptic Painlev\'e
equation.
As a byproduct, we find that ${}_{10}E_9$ has 
the affine Weyl group symmetry of type $E^{(1)}_7$.

\section{Geometric formulation}
In this section, we give a geometric interpretation of 
the results in the previous section. 
\par\medskip

When the nine points $P_1, \ldots, P_9$ are in special positions such that
they are the base of an elliptic fibration $X \rightarrow \BP^1$,
the translation subgroup $\BZ^8 \subset W(E^{(1)}_8)$ is realized as the 
Mordell-Weil group of the elliptic fibration. This fact follows from 
\cite{M},Theorem 5, 6. 
In the case of the Painlev\'e equation, the configuration of the points
$P_1,\ldots, P_9$ is not special in the sense above. However,
for the translations of the type $T_{\alpha_{ij}}$, one can also find the
special configuration at the intermediate step of the translation, and 
the above correspondence between the translations and Mordell-Weil 
group is applicable also for the Painlev\'e equation.

Let us again consider the translation $T_{\alpha_6}$ as an example.
Under the translation $T_{\alpha_6}$, the points $P_i$ ($i \neq 6,7,10$)
are invariant and the new points $\overline{P}_6$ and $\overline{P}_7$ are 
determined so that
\begin{equation}\label{P8move}
P_6+P_7=\overline{P}_6+\overline{P}_7, \quad
P_1+\cdots+P_6+\overline{P}_7+P_8+P_9=0,
\end{equation}
with respect to the addition on the cubic $C_0$.
This means that $\overline{P}_7$ is the additional intersection 
point of the elliptic 
pencil defined by the eight points $P_i$ ($i \neq 7,10$). 
Using this pencil of cubics, the transformation $T_{\alpha_6}(P_{10})$
is geometrically described as follows.
Consider a cubic curve $C$ passing through the nine points
$P_i$ ($i \neq 7$). The new point $\overline{P}_{10}$ is determined by 
\begin{equation}
\overline{P}_{10}+\overline{P}_{7}=P_{10}+P_6,
\end{equation}
with respect to the addition on the curve $C$. 
In this sense, the elliptic 
Painlev\'e equation is a non-autonomous deformation of the addition formula of
the elliptic function, where the addition is defined on the moving curve $C$.

Let us apply this formulation to construct the hypergeometric solutions.
To do this, consider the case when the three points $P_4,P_5,P_9$ are on a 
line $\ll$. In such a case, 
if $P_{10} \in \ll$ then $\overline{P}_{10} \in \ll$, 
since the curve $C$ is decomposed into the line $\ll$ 
(passing through $P_4, P_5, P_9, \overline{P}_{10}$) 
and the conic $C_2$ (passing through $P_1, P_2, P_3, P_6, \overline{P}_7, P_8$).
Let us put
\begin{equation}
x=P_6, \quad \oy=\overline{P}_7,
\quad f=P_{10}, \quad \of=\overline{P}_{10},
\end{equation}
then these points are in the configuration shown in Figure.1.

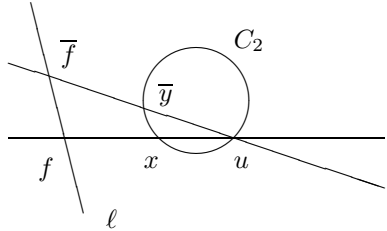
\begin{figure}[h]
\setlength{\unitlength}{1mm}
\begin{picture}(60,25)(-20,7)
\put(40,20){\circle{20}}
\put(15,15){\line(3,0){50}}
\put(15,25){\line(3,-1){50}}
\put(25,5){\line(-1,4){7}}
\put(19,10){$f$}
\put(22,25){$\of$}
\put(45,11){$u$}
\put(33,11){$x$}
\put(35,20){$\oy$}
\put(45,27){$C_2$}
\put(28,3){$\ll$}
\end{picture}
\caption{Configuration of the points}
\end{figure}
By elementary geometry we have
the linear difference equation for $f$:
\begin{lem}
\begin{equation}\label{fg-rel1}
\begin{array}c
(a,x)\of=(a,\oy) D f-(a,D f)\oy, \\[3mm]
(a,\oy)f=(a,x)D^{-1}\of-(a,D^{-1}\of)x,
\end{array}
\end{equation}
where $D={\rm diag}\left(\dfrac{x_2x_3}{\oy_2\oy_3},\dfrac{x_3x_1}{\oy_3\oy_1},
\dfrac{x_1x_2}{\oy_1\oy_2} \right)$, $(a,x)=a_1 x_1+a_2 x_2+a_3 x_3=0$ is 
the equation of the line $\ll$, 
and $x_i$, $\oy_i$ $(i=1,2,3)$ are
homogeneous coordinates for $x$ and $\oy$, respectively.
\end{lem}
The normalization of the variables
$f$ is chosen so that $f_i=\tau(s_0.\E_i)/\tau(\E_i)$.
%
{}From eq.(\ref{fg-rel1}), it is easy to deduce the following:

\begin{prop}
\begin{equation}\label{3term-f}
\dfrac{y_1\oy_3 (a,x)[\oy_2 \of_1-x_2 f_1]}{(x_2\oy_3-x_3\oy_2)}-
\dfrac{x_1\ux_3 (a,y)[\ux_2 \uf_1-y_2 f_1]}{(\ux_2y_3-\ux_3y_2)}=
a_2(x_1y_2-x_2y_1)f_1.
\end{equation}
\end{prop}

Using the parameterization of the points $P_i$ ($1 \leq i \leq 9$)
by the theta function 
$P_i=\left(\dfrac{[23i]}{[1i]}:\dfrac{[13i]}{[2i]}:
\dfrac{[12i]}{[3i]} \right)$, 
we finally obtain the following:

\begin{thm} When the points $P_4, P_5, P_9$ and $P_{10}$
are on a line, the elliptic
Painlev\'e equation for the translation $T_{\alpha_6}$ is reduced to the
linear difference equation
\begin{equation}\label{3term-theta}
c_1\left(\dfrac{[26][13{\overline 7}]}{[2{\overline 7}][136]}
\of_1-f_1\right)
-c_2\left(\dfrac{[27][13{\underline 6}]}{[2{\underline 6}][137]}
\uf_1-f_1\right)
=c_3f_1,
\end{equation}
where 
\begin{equation}
\begin{array}l
c_1=[27][2{\overline 7}][46][56][12{\overline 7}]
[136][237][456]/[6{\overline 7}],\\[3mm]
c_2=[26][2{\underline 6}][47][57][12{\underline 6}]
[137][236][457]/[7{\underline 6}],\\[3mm]
c_3=[24][25][76][123][1{\overline 7}6][245][376],\\[3mm]
\ep_{\overline{7}}=\ep_7+\delta, \quad 
\ep_{\underline{6}}=\ep_6+\delta.
\end{array}
\end{equation}
\end{thm}
By suitable gauge transformation and parameter change, 
eq.(\ref{3term-theta}) also coincides with eq.(\ref{10E9-eq}).

The above geometric formulation can be also applied for the degenerate cases.
In particular, when the cubic $C_0$ factors into a conic and a line
[resp. three lines], then
the symmetry reduces to $W(E^{(1)}_7)$ [resp. $W(E^{(1)}_6)$]
and the corresponding hypergeometric solution
is reduced to the Askey-Wilson [resp. big $q$-Jacobi] functions.

We believe that our results will open a new avenue to the
systematic study of the special functions (Askey scheme) from the point 
of view of the affine Weyl group symmetry.

\medskip
\noindent
{\it Acknowledgments}

We would like to thank M.-H.~Saito, V.P. Spiridonov and A.~S.~Zhedanov
for valuable discussions.

\end{document}